\documentclass[journal=jacsat,manuscript=article]{achemso}
\usepackage[version=3]{mhchem}
\usepackage{graphicx}
\usepackage{amsmath}
\usepackage{newtxmath}
\usepackage{textcomp}
\usepackage{xcolor}
\usepackage{soul}
\usepackage{colortbl}
\usepackage{units}
\usepackage{caption}
\usepackage{float}
\usepackage[normalem]{ulem}
\usepackage{verbatim}
\usepackage{lineno}
\usepackage{comment}
\usepackage{bm}
\makeatletter
\let\saved@includegraphics\includegraphics
\AtBeginDocument{\let\includegraphics\saved@includegraphics}
\renewenvironment*{figure}{\@float{figure}}{\end@float}
\makeatother

\newcommand{\reftextit}[1]{}

\DeclareCaptionLabelFormat{adja-page}{\hrulefill\\#1 #2 \emph{(previous page)}}

\title[An \textsf{achemso} demo]{Selective Excitation of Bloch Modes in Canalized Polaritonic Crystals}
\let\oldalign\align
\let\oldendalign\endalign
\renewenvironment{align}
  {\linenomathNonumbers\oldalign}
  {\oldendalign\endlinenomath}
\let\oldequation\equation
\let\oldendequation\endequation
\renewenvironment{equation}
  {\linenomathNonumbers\oldequation}
  {\oldendequation\endlinenomath}
\clearpage
\author{\normalsize Yanzhen Yin}
\affiliation[Tongji University]
{%
\footnotesize MOE Key Laboratory of Advanced Micro-Structured Materials, Shanghai Frontiers Science Center of Digital Optics, Institute of Precision Optical Engineering, and School of Physics Science and Engineering, Tongji University, Shanghai 200092, China}
\altaffiliation{Contributed equally to this work}

\author{Zhichen Zhao}
\affiliation[Tongji University]
{%
\footnotesize MOE Key Laboratory of Advanced Micro-Structured Materials, Shanghai Frontiers Science Center of Digital Optics, Institute of Precision Optical Engineering, and School of Physics Science and Engineering, Tongji University, Shanghai 200092, China}
\altaffiliation{Contributed equally to this work}

\author{Junbo Xu}
\affiliation[Tongji University]
{%
\footnotesize MOE Key Laboratory of Advanced Micro-Structured Materials, Shanghai Frontiers Science Center of Digital Optics, Institute of Precision Optical Engineering, and School of Physics Science and Engineering, Tongji University, Shanghai 200092, China}

\author{Zerui Wang}
\affiliation[Tongji University]
{%
\footnotesize MOE Key Laboratory of Advanced Micro-Structured Materials, Shanghai Frontiers Science Center of Digital Optics, Institute of Precision Optical Engineering, and School of Physics Science and Engineering, Tongji University, Shanghai 200092, China}

\author{Lei Zhou}
\affiliation[Tongji University]
{%
\footnotesize MOE Key Laboratory of Advanced Micro-Structured Materials, Shanghai Frontiers Science Center of Digital Optics, Institute of Precision Optical Engineering, and School of Physics Science and Engineering, Tongji University, Shanghai 200092, China}

\author{Zhou Zhou}
\affiliation[Tongji University]
{%
\footnotesize MOE Key Laboratory of Advanced Micro-Structured Materials, Shanghai Frontiers Science Center of Digital Optics, Institute of Precision Optical Engineering, and School of Physics Science and Engineering, Tongji University, Shanghai 200092, China}

\author{Yu Yin}
\affiliation[Tongji University]
{%
\footnotesize MOE Key Laboratory of Advanced Micro-Structured Materials, Shanghai Frontiers Science Center of Digital Optics, Institute of Precision Optical Engineering, and School of Physics Science and Engineering, Tongji University, Shanghai 200092, China}

\author{Di Huang}
\affiliation[Tongji University]
{%
\footnotesize MOE Key Laboratory of Advanced Micro-Structured Materials, Shanghai Frontiers Science Center of Digital Optics, Institute of Precision Optical Engineering, and School of Physics Science and Engineering, Tongji University, Shanghai 200092, China}

\author{Gang Zhong}
\affiliation[Macau University of Science and Technology]
{%
\footnotesize Macao Institute of Materials Science and Engineering (MIMSE), Faculty of Innovation Engineering, Macau University of Science and Technology, Taipa, Macao, 999078, China}

\author{Xiang Ni}
\affiliation[Central South University]
{%
\footnotesize School of Physics, Central South University, Changsha, Hunan, 410083, China}

\author{Zhanshan Wang}
\affiliation[Tongji University]
{%
\footnotesize MOE Key Laboratory of Advanced Micro-Structured Materials, Shanghai Frontiers Science Center of Digital Optics, Institute of Precision Optical Engineering, and School of Physics Science and Engineering, Tongji University, Shanghai 200092, China}
\alsoaffiliation[Tongji University]
{%
\footnotesize Shanghai Institute of Intelligent Science and Technology, Tongji University, Shanghai 200092, China}

\author{Xinbin Cheng}
\affiliation[Tongji University]
{%
\footnotesize MOE Key Laboratory of Advanced Micro-Structured Materials, Shanghai Frontiers Science Center of Digital Optics, Institute of Precision Optical Engineering, and School of Physics Science and Engineering, Tongji University, Shanghai 200092, China}
\alsoaffiliation[Tongji University]
{%
\footnotesize Shanghai Institute of Intelligent Science and Technology, Tongji University, Shanghai 200092, China}

\author{Jingyuan Zhu}
\affiliation[Tongji University]
{%
\footnotesize MOE Key Laboratory of Advanced Micro-Structured Materials, Shanghai Frontiers Science Center of Digital Optics, Institute of Precision Optical Engineering, and School of Physics Science and Engineering, Tongji University, Shanghai 200092, China}
\email{zjy_tongji@tongji.edu.cn}

\author{Qingdong Ou}
\affiliation[Macau University of Science and Technology]
{%
\footnotesize Macao Institute of Materials Science and Engineering (MIMSE), Faculty of Innovation Engineering, Macau University of Science and Technology, Taipa, Macao, 999078, China}
\email{qdou@must.edu.mo}

\author{Tao Jiang}
\affiliation[Tongji University]
{%
\footnotesize MOE Key Laboratory of Advanced Micro-Structured Materials, Shanghai Frontiers Science Center of Digital Optics, Institute of Precision Optical Engineering, and School of Physics Science and Engineering, Tongji University, Shanghai 200092, China}
\alsoaffiliation[Tongji University]
{%
\footnotesize Shanghai Institute of Intelligent Science and Technology, Tongji University, Shanghai 200092, China}
\email{tjiang@tongji.edu.cn}
\begin{document}

\maketitle
{\renewcommand{\baselinestretch}{1.7}
\clearpage
\begin{abstract}
Polaritonic crystals (PoCs) have experienced significant advancements through involving hyperbolic polaritons in anisotropic materials such as $\upalpha$-MoO$_{\rm 3}$, offering a promising approach for nanoscale light control and improved light-matter interactions.
Notably, twisted bilayer $\upalpha$-MoO$_{\rm 3}$ enables tunable iso-frequency contours (IFCs), especially generating flat IFCs at certain twist angles, which could enhance mode selectivity in their PoCs through the highly collimated and canalized polaritons.
This study unveils the selective excitation of Bloch modes in PoCs with square-lattice structures on twisted bilayer $\upalpha$-MoO$_{\rm 3}$ with canalized phonon polaritons.
Through the optimization of the square lattice design, there is an effective redistribution of canalized polaritons into the reciprocal lattices of PoCs. 
Fine-tuning the periodicity and orientation of the hole lattice enables momentum matching between flat IFCs and co-linear reciprocal points, allowing precise and directional control over desired Bragg resonances and Bloch modes.
This research establishes a versatile platform for tunable polaritonic devices and paves the way for advanced photonic applications.

\end{abstract}
}
\maketitle
\renewcommand{\baselinestretch}{2}
\setlength{\parskip}{7pt}

\newpage
\section{Introduction}
Polaritons, as half-light half-matter quasiparticles formed by the strong coupling of photons with material excitations, offer a unique ability to confine and control light at sub-wavelength scales\cite{2014-Basov-Science-Tunable-Phonon-Polaritons-in-Atomically,2015-Basov-NatNano-Graphene-on-Hexagonal-Boron,2015-Dickson-A.M.-Hyperbolic,2016-Abajo-Science-Polaritons-in-van,2017-Koppens-N.M.-Polaritons-in-layered,2018-QiaoliangBao-Nature-In-Plane-Anisotropic-and-Ultra-Low,2020-Alu-AOM-Phonon-polaritons,2020-P.Alonso-Gonzalez-Nat.Mat.-Broad-spectral-tuning-of-ultra-low-loss-polaritons,2021-P.N.Li-Nature-Ghost-hyperbolic,2022-QingDai-NatNano-Doping-Driven-Topological,2023-DaiQing-Science-Gate-tunable-negative-refraction-of-mid-infrared-polaritons,chen2012optical,mancini2024multiplication,lu2024tailoring}.
This advantage has led to the development of polaritonic crystals (PoCs), which are polaritonic materials with periodic structures that unlock new possibilities for manipulating light-matter interactions at the nanoscale\cite{alfaro2019deeply,alfaro2021hyperspectral,capote2022twisted,sahoo2023polaritons,lv2023hyperbolic,khanikaev2013photonic}.
Through tailored design, PoCs serve as a versatile platform for effectively exciting and controlling Bloch modes and Bragg resonances, which emerge from the additional momentum components provided by the periodic structures\cite{sahoo2023polaritons,lv2023hyperbolic}.
This progress fosters technological innovations in cutting-edge applications such as nanoimaging\cite{chen2019modern,yao2020nanoimaging}, optical sensing\cite{li2018boron,zhang2021interface}, and on-chip optical communications and computations\cite{jiang2014metal,wang2024planar}.

Various polaritonic materials have been employed to realize PoCs, including graphene\cite{sunku2018photonic,xiong2019photonic,luo2020situ,hesp2021nano}, which supports surface plasmon polaritons (SPPs), hexagonal boron nitride (hBN)\cite{alfaro2019deeply,alfaro2021hyperspectral,orsini2024deep,golz2024revealing}, which supports in-plane isotropic phonon polaritons (PhPs), and $\upalpha$-MoO$_3$\cite{capote2022twisted,huang2023plane,sahoo2023polaritons,lv2023hyperbolic}, which supports in-plane hyperbolic PhPs.
Compared to the in-plane isotropic polaritons found in graphene or hBN, the hyperbolic PhPs in $\upalpha$-MoO$_3$ provide directional control of Bloch modes, attributed to their anisotropic iso-frequency contours (IFCs) and unique propagation characteristics\cite{lv2023hyperbolic}.
However, the complex distribution of hyperbolic IFCs in momentum space often results in the excitation of Bragg resonances with unwanted resonance orders, hindering the precise excitation of Bloch modes at specific momenta and directions.
Recent progress in twist optics has highlighted the twisted bilayer $\upalpha$-MoO$_3$ system as a promising approach for manipulating the IFCs of PhPs\cite{2020-Andrea-Nature-Topological-Polaritons-and-Photonic-Magic,2020-SiyuanDai-NatMat-Configurable-Phonon-Polaritons,2020-ShaozhiDeng-Nanolett-Phonon-Polaritons-in-Twisted,2020-Alu-AOM-Phonon-polaritons,2020-Pablo-Nanolett-Twisted-Nano-Optics-Manipulating,zhou2023gate,oudich2024engineered,duan2023multiple}.
By carefully engineering the stacking of $\upalpha$-MoO$_3$ bilayers, incorporating specific thicknesses and twist angles, canalized polariton at designated excitation wavelengths will emerge\cite{2020-Andrea-Nature-Topological-Polaritons-and-Photonic-Magic}.
Such polaritons are distinguished by their diffractionless and highly collimated propagation characteristics\cite{2020-Rainer-NatCommu-Collective-Nearfield-Coupling,2020-Andrea-Nature-Topological-Polaritons-and-Photonic-Magic,duan2023multiple}, exhibiting quasi-linear, asymmetric, and highly localized IFCs.
By fine-tuning the periodicity and orientation of the square lattice, a precise alignment can be achieved between these flat IFCs in twisted bilayer $\upalpha$-MoO$_3$ and the reciprocal lattices of PoCs, thereby enhancing their coupling.
This enhancement enables the selective excitation of specific Bragg resonances and Bloch modes.

In this work, we fabricate PoCs using twisted bilayer $\upalpha$-MoO$_3$, demonstrating the selective excitation of Bloch modes through the coupling between flat IFCs and reciprocal lattice of PoCs via near-field nanoimaging, and analyzing the reflection coefficient of Bragg resonances across various structures.
By adjusting the periodicity of the lattice structures while maintaining the momentum matching between reciprocal lattice and flat IFCs, we successfully achieve canalized Bloch modes with momenta corresponding to the designed Bragg resonance orders.
We further demonstrate the selective excitation of arbitrary Bragg resonance orders in PoCs by varying the orientation of lattice structures relative to the crystallographic axes and observe highly collimated polariton propagation.
Moreover, we investigate the dependence of the reflection coefficient on lattice periodicity and orientation of PoCs, highlighting the tunability of selective excitation in the far field.
Our work realizes a new platform for precisely configurable Bloch modes in canalized PoCs with flat IFCs, opening avenues for advanced applications in on-chip polaritonic device engineering and optical communication technologies.

\section{Results}
In our study, as illustrated in Figure 1a, the PoCs feature a series of square-type hole arrays etched in the twisted bilayer $\upalpha$-MoO$_{\rm 3}$ on the SiO$_2$ substrate.
For the twisted bilayer $\upalpha$-MoO$_{\rm 3}$, we sequentially stack two $\upalpha$-MoO$_{\rm 3}$ layers onto the substrate at appropriate twist angle and thickness determined in Supporting Information (SI) section 1 to support flat IFCs at our experimental wavenumber of 931 cm$^{-1}$.
The flat IFC also provides the direction of polariton canalization, allowing us to define a Cartesian coordinate for each sample in which the $x$ and $y$ axes are parallel and perpendicular to the direction of canalization.
We further acquire momentum space maps for unperforated samples using transfer matrix method (TMM; see details in SI section 2) to validate our designed parameters.
As depicted in Figure 1b and Figure S4, the momentum space maps not only demonstrate expected quasi-linear dispersions almost parallel to $k_y$ axis, but also present asymmetric and highly localized distributions that facilitates the precise excitation of resonance orders.

\begin{figure}[htbp]
    \centering
    \includegraphics[width=16cm]{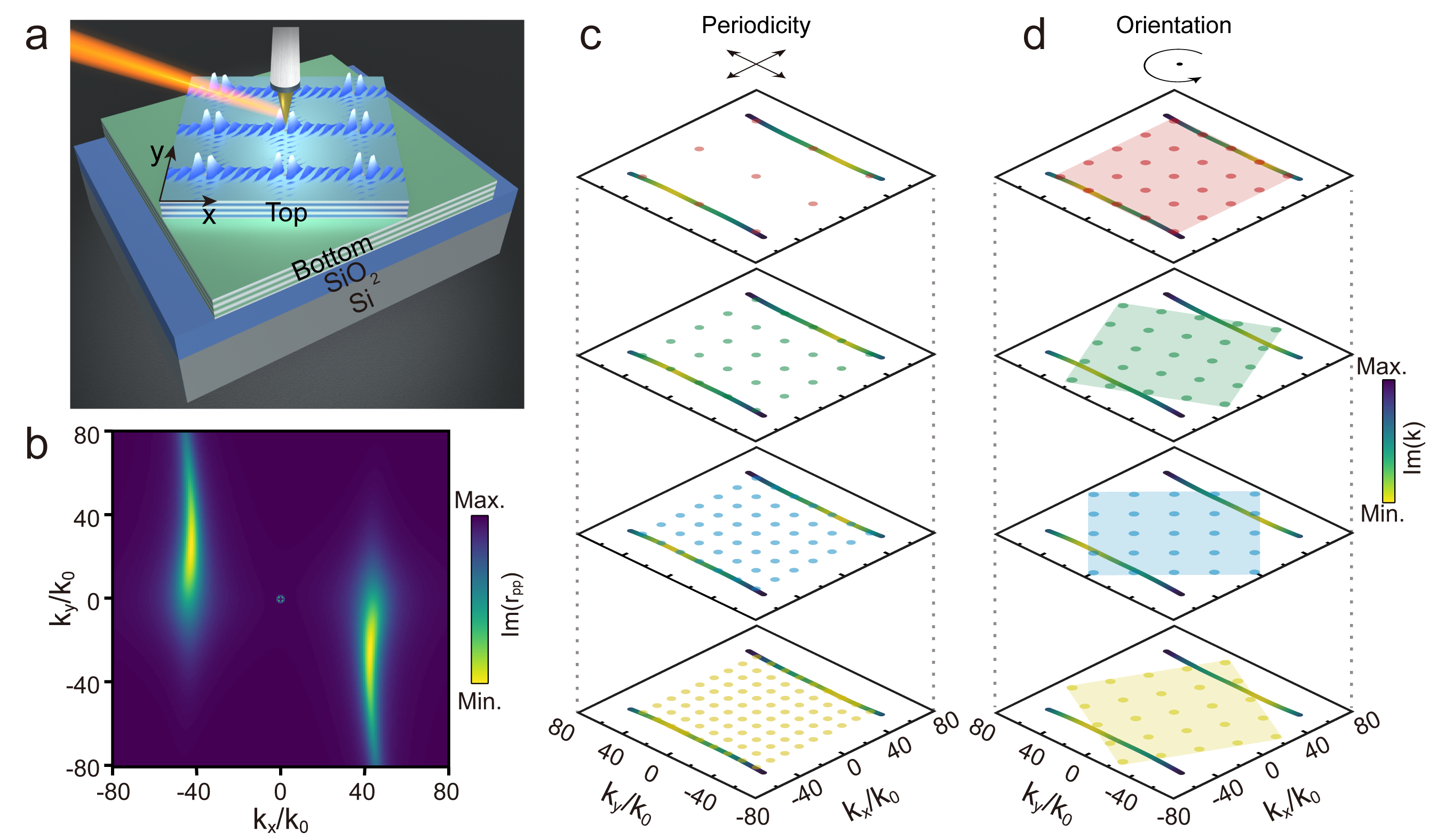}
    \caption{Canalized PoC structure design based on twisted bilayer $\upalpha$-MoO$_{\rm 3}$.
    a) Schematic of IR \textit{s}-SNOM measurement for the canalized PoC with square lattice in twisted bilayer $\upalpha$-MoO$_{\rm 3}$.
    b) Momentum space map of twisted bilayer $\upalpha$-MoO$_{\rm 3}$ possessing polariton canalization calculated by TMM.
    c) Modulating the lattice periodicity $P$ and d) the angle $\theta$ between lattice orientation and the orientation of flat IFCs the while maintaining the momentum matching between flat IFCs and the co-linear reciprocal points.
    }
    \label{fig:fig1}
\end{figure}

On the basis of the canalized polaritons in the twisted bilayer $\upalpha$-MoO$_{\rm 3}$ above, we construct PoCs with square-lattice hole arrays, leading to a square reciprocal lattice as a consequence.
As the excitation of Bloch modes is fundamentally governed by the momentum matching between the polaritonic dispersions and the reciprocal lattice\cite{sahoo2023polaritons,lv2023hyperbolic}, where the flat IFCs enable the precise excitation of any resonance order by fine-tuning the lattice structure. 
To unveil the precise tunablity of configurable Bloch modes in canalized PoCs, we developed two series of PoCs.
These series respectively exploring the modulation effect of the lattice periodicity $P$ (as shown in Figure 1c) and the orientation angle $\theta$ (depicted in Figure 1d). The orientation angle $\theta$ is defined as the angle between the reciprocal lattice and the flat IFCs.
Particularly, all the PoCs are etched with circular holes with a uniform diameter of $\unit[50]{nm}$.
Considering the measured wavelengths of PhPs are half of their actual values as we employ scattering near-field optical microscopy ($s$-SNOM) for interference measurements (see details in experiment section), we double the value of calculated IFCs and align them with co-linear reciprocal lattice during modifying the lattice periodicities and orientations as depicted in Figure 1c and 1d to facilitate the observation of Bloch modes.

For investigation of the $P$ modulation, we etch hole arrays in the twisted bilayer $\upalpha$-MoO$_{\rm 3}$ of which the flat IFC is depicted in Figure 1b.
The design features the top and bottom layers with thicknesses of $\unit[64]{nm}$ and $\unit[87]{nm}$, respectively, stacked together with a twist angle of $62^{\circ}$.
According to Figure S5a-d, hole arrays with different $P$ of $\unit[115]{nm}$, $\unit[230]{nm}$, $\unit[345]{nm}$, and $\unit[460]{nm}$ are etched to ensure momentum matching at specific lattice orders (indicated by $\pm n$ along $k_x$ direction, $\pm m$ along $k_y$ direction, where $n$ and $m$ are nonnegative integers).
Following the fabrication of PoCs, we conduct \textit{s}-SNOM measurements and perform theoretical calculations using Rigorous Coupled Wave Analysis (RCWA)\cite{2011-C.Raymond-Prog.Electromagn.Res.B-Improved-formulation-of-scattering-matrices} (SI section 3 for details).
Figures 2a-d depict the experimental near-field images of PoCs with different $P$, presenting canalized PhPs interference patterns that are predominantly along the $x$ axis and almost absent along the $y$ axis.
As $P$ increases, the distinct bright fringes between two adjacent holes along the $x$ axis exhibit an increase in number from one to four, which corresponds to the alignment between flat IFC and resonance orders with $n = \pm 1, 2, 3, 4$, indicating the excitation of different orders of Bloch modes.
The calculated near-field patterns (Figure 2e-h) agree well with the corresponding experimental results, confirming the highly directional excitation of Bloch modes.

\begin{figure}[htbp]
    \centering
    \includegraphics[width=16cm]{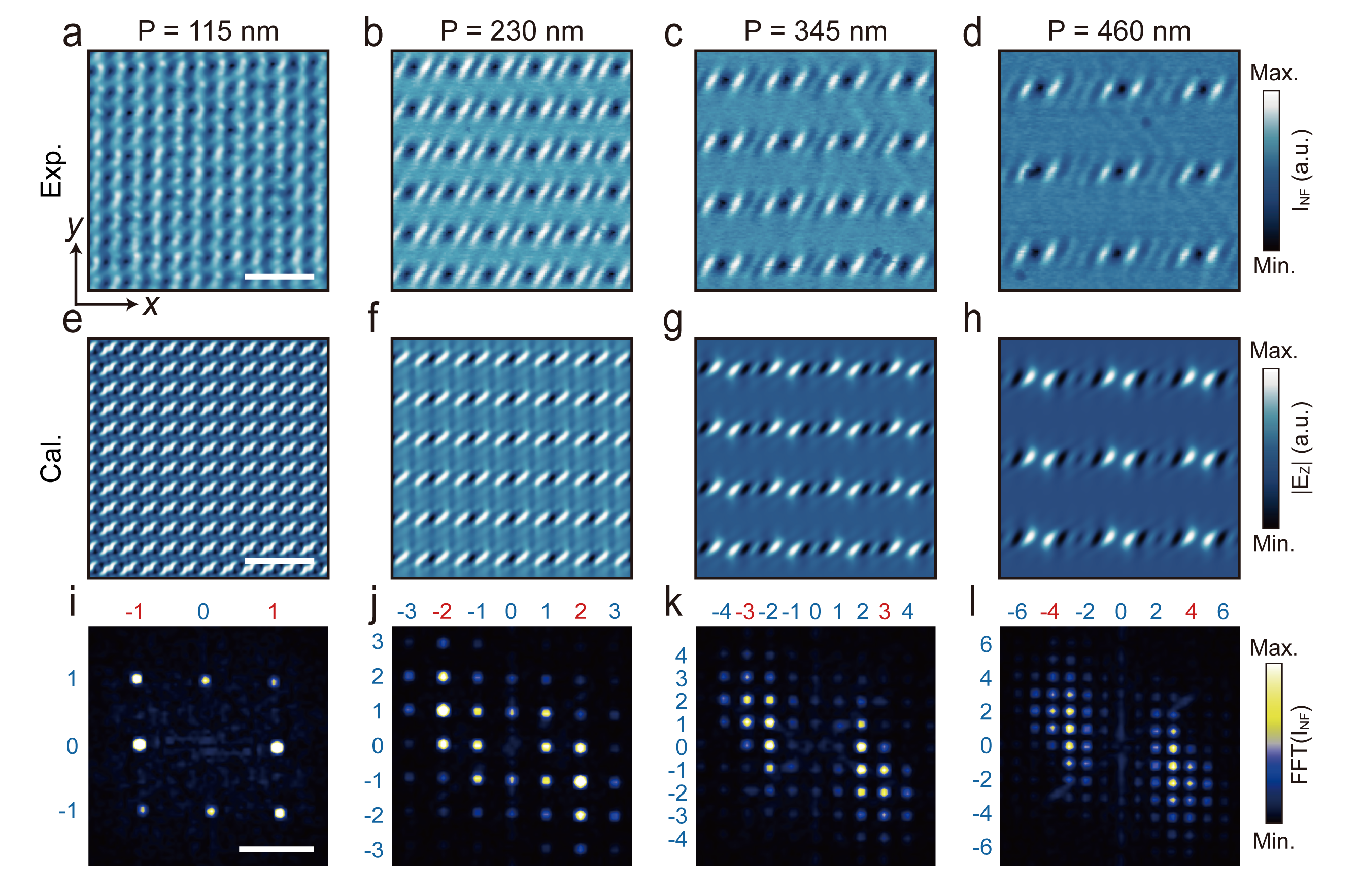}
    \caption{Lattice periodicity modulation of canalized PoCs.
    a-d) Near-field images, e-h) calculated field distributions and i-l) FFT amplitude maps of the measured near-field images at different periodicity ($P$) of $\unit[115]{nm}$, $\unit[230]{nm}$, $\unit[345]{nm}$ and $\unit[460]{nm}$, respectively.
    The $x$ axis in panel (a) is aligned with the propagation direction of the canalized PhPs.
    The red numbers define the different resonance orders.
    Scale bar in panels (a-h): $\unit[400]{nm}$.
    Scale bar in panels (i-l): 50$k_0$.
    }
    \label{fig:fig2}
\end{figure}

By processing the near-field images from both experiments and calculations with Fast Fourier Transform (FFT), we can directly determine the resonance orders of Bloch modes in momentum space as illustrated in Figure 2i-l and Figure S5e-h.
Notably, the FFT amplitude exhibit an asymmetric and discrete distribution, inheriting both the characteristics of the square reciprocal lattice of PoCs and the asymmetric quasi-linear IFC of the PhPs in twisted bilayer $\upalpha$-MoO$_{\rm 3}$.
This analysis highlights the configurable excitation of various resonance orders across different $P$ values in both experimental and theoretical frameworks.
While the results show good agreements between experiments and calculations with $P$ of $\unit[115]{nm}$ and $\unit[230]{nm}$, Figure 2k and 2l show some deviation from the corresponding simulated results (Figure S5g and S5h) with the main excited resonance orders ($\pm 2$, $\pm m$) and ($\pm 3$, $\pm m$).
We attribute this mismatch to the loss of PhPs during propagation between neighboring holes with greater distances of $\unit[345]{nm}$  and $\unit[460]{nm}$ in the experiment.
Moreover, as $P$ increases, there emerges a progressive transition in the amplitude from a discretedistribution to a continuum-like distribution, closely resembling Fiqure 1b. This is attributed to theincreased distance between the nearest holes, which diminishes the modulation effect of theperiodic lattice.

Next, we turn our attention to the modulation effect of $\theta$ based on another twisted bilayer $\upalpha$-MoO$_{\rm 3}$, of which the IFC exhibits similar linearity, asymmetry and localization (Figure S4). This system comprises top and bottom layers with thicknesses of $\unit[65]{nm}$ and $\unit[96]{nm}$, respectively. The twist angle is set at $61^{\circ}$.
With respect to the direction of canalized polaritons, the hole arrays are etched at $\theta$ of $-5^{\circ}$, $21.6^{\circ}$, $40^{\circ}$, and $58.3^{\circ}$, respectively.
As shown in Figure 3, at $\theta = -5^{\circ}$, the alignment of reciprocal lattice at orders ($\pm 2$, $\pm m$) with the flat IFCs results in two bright fringes between adjacent holes along the $x$ axis (Figure 3a).
While for other $\theta$ of $21.6^{\circ}$, $40^{\circ}$, and $58.3^{\circ}$, the bright fringes are not confined between the two nearest holes but still align with the direction of the $x$ axis, demonstrating a manifestation of polariton canalization.
This behavior highlights the strong directional robustness of the system, indicating that the canalization effect maintains the directional propagation of polaritons irrespective of the specific spatial arrangement of the fringes.
Our simulations (Figure 3e-h) show sharp fringes that align well with the experimental results.

\begin{figure}[htbp]
    \centering
    \includegraphics[width=16cm]{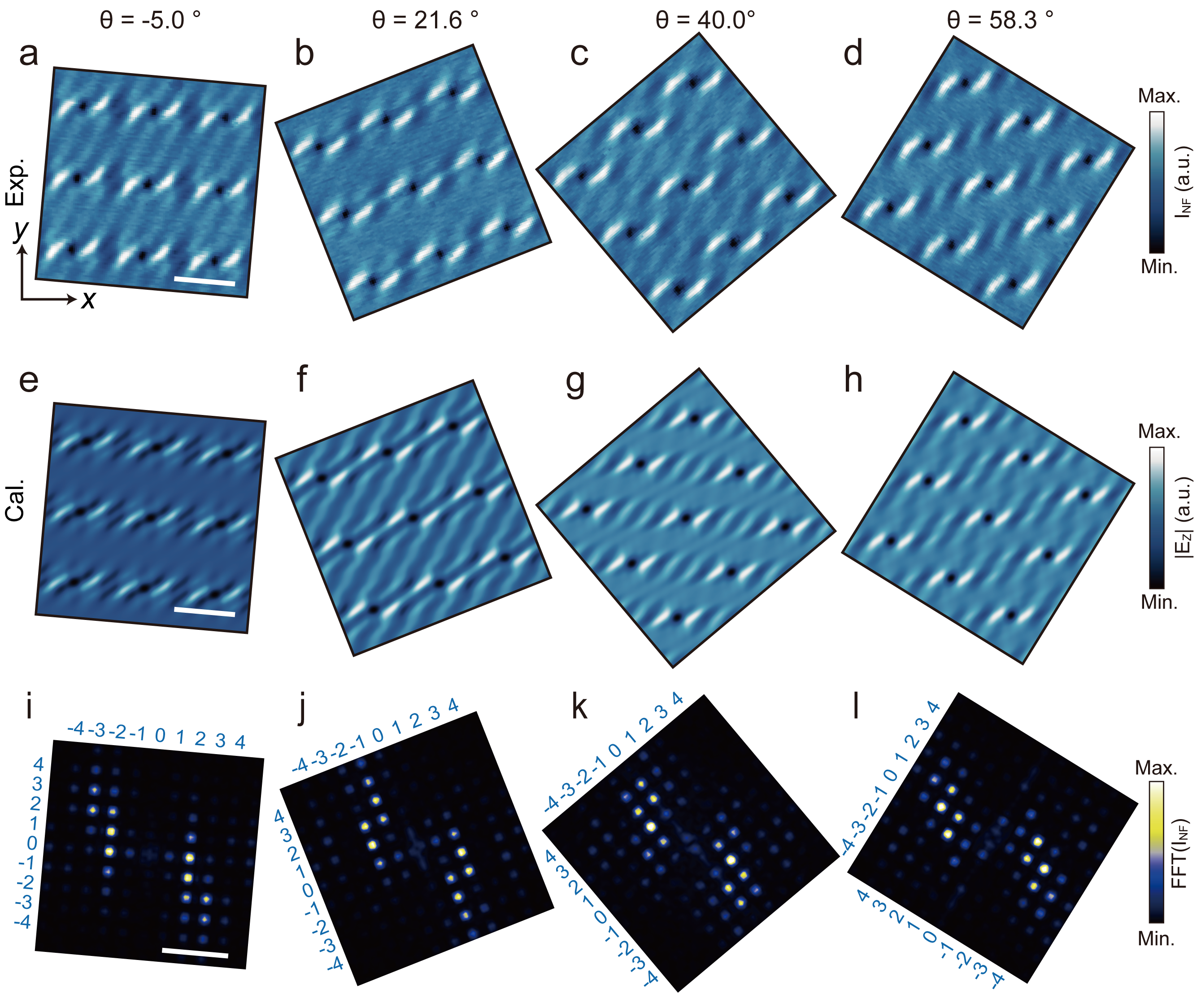}
    \caption{Lattice orientation modulation of canalized PoCs.
    a-d) Near-field images, e-h) calculated field distributions and i-l) FFT amplitude maps of the measured near-field images at different angles $\theta$ between lattice orientation and collimated direction of PhPs ($\theta$) of $-5^{\circ}$, $21.6^{\circ}$, $40^{\circ}$ and $58.3^{\circ}$, respectively.
    The $x$ axis in panel (a) is aligned with the propagation direction of the canalized PhPs.
    Scale bar in panels (a-h): $\unit[300]{nm}$.
    Scale bar in panels (i-l): 50$k_0$.
    }
    \label{fig:fig3}
\end{figure}

The FFT amplitude maps of experimental results demonstrate that the change in $\theta$ mainly affects the resonance orders. as shown in Figure 3i-l,
With different $\theta$, the reciprocal lattice rotates accordingly, allowing the flat IFC to align with different co-linear resonance orders that are not accessible at $\theta = -5^{\circ}$.
Simultaneously, the excited resonance orders are concentrated asymmetrically at reciprocal space points, which exhibit stronger intensities in the momentum space map of unperforated twisted bilayer $\upalpha$-MoO$_{\rm 3}$ according to the asymmetric IFC (Figure S4).
Such variation in the selection of Bragg resonance orders can also be interpreted from the FFTs obtained from calculations as depicted in Figure S6e-h.
Therefore, the precise excitation of arbitrary order Bragg resonance in canalized PoCs is available via changing the lattice orientation to arrange the alignment in momentum space.

Beyond examining the near-field nanoimaging resulting from Bloch modes in PoCs, our study extends to a comprehensive investigation of the infrared optical responses in these materials. 
This effort involves analyzing the interplay between the parameters of PoCs and their reflection coefficients, which are calculated by utilizing the RCWA method.
Our analysis is based on the PoCs with a periodicity of $\unit[230]{nm}$, where the flat IFCs approximately match with the momenta of the (1,-1), (1,0), and (1,1) Bragg resonance orders.
As we increase $P$ by a factor of $n$, where $n$ is a positive integer, intervals between adjacent reciprocal lattice points are simultaneously reduced by $n$ times, as illustrated in Figure 2.
This operation results in a denser overlap and consequently modified interactions between the flat IFCs and the reciprocal lattices as depicted in Figure 4a.
Specifically, it marks a transition from the limited intersection between the flat IFCs and the reciprocal lattices from ($n$,$-n$) to ($n$,$n$) orders.
To clarify the impact arising from periodicity, we calculate the reflection coefficient spectra corresponding to Bragg resonance orders at ($n$,$-n$), ($n$,$0$), and ($n$,$n$) by tweaking $n$ under normal incidence, with incident light field aligning along the direction of polariton canalization.
The calculated results are further elaborated in Figure S7.
The differences of reflection coefficients among the three orders with the same $n$ primarily arise from the intrinsic asymmetry of the flat IFCs in twisted bilayer $\upalpha$-MoO$_{\rm 3}$, as depicted Figure 4b.
When increasing $n$, we observe a decrease in the reflection coefficients associated with the Bragg resonances, despite their identical positions in momentum space. 
The reduction can be attributed to high-order diffracted waves undergoing more scattering processes, resulting in a lower diffraction amplitude.
We further note the intricate interplay between intrinsic polariton dispersion and the periodic structures in PoCs, hinted by the distinct spectral responses of the reflection coefficients with different $n$ , as depicted in Figure S7c.

\begin{figure}[htbp]
    \centering
    \includegraphics[width=16cm]{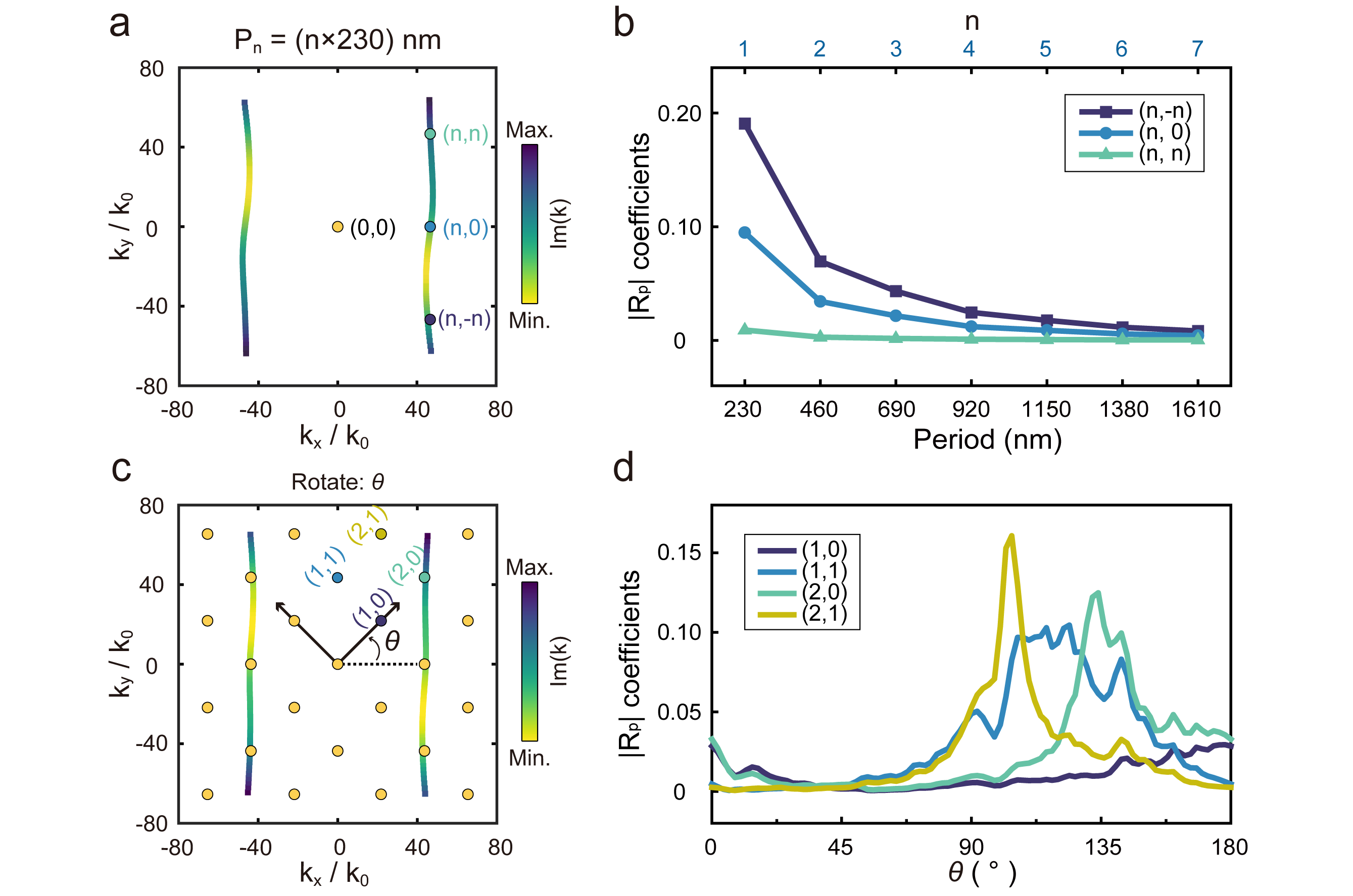}
    \caption{Tunable reflection coefficients in canalized PoCs.
    (a) Momentum-space distribution including IFCs of canalized PoCs and Bragg resonance orders, calculated for the lattice periodicity designed to match the momentum of the IFCs.
    Notably, due to the absence of near-field interference measurements of PhPs, the calculated IFC values do not need to be doubled as in the previous figures.
    Thus a periodicity of $P = \unit[230]{nm}$ corresponds to the $n = 1$ orders.
    (b) Periodicity-dependent reflection coefficient curves corresponding to the ($n$, $-n$), ($n$, 0), and ($n$, $n$) orders.
    (c) Momentum-space distribution including IFCs of canalized PoCs and Bragg resonance orders, calculated for $\unit[348]{nm}$ periodicity and $45^{\circ}$ rotation angle.
    (d) Orientation-dependent reflection coefficient curves corresponding to the (1,0), (1,1), (2,0), and (2,1) orders.
    }
    \label{fig:fig4}
\end{figure}

By progressively rotating the square lattice, the Bragg resonance orders, which rotate simultaneously, intersect with the flat IFCs at specific orientation angles $\theta$. This process enables selective excitation of Bragg resonance orders, as presented in Figure 4c.
We investigate the reflection coefficient against the orientation angle $\theta$, spaning from $0^{\circ}$ to $180^{\circ}$ considering the inversion symmetric IFCs and reciprocal lattices.
With an adjustable $\theta$, the collection of reachable points by rotating $\theta$ corresponding to the ($\pm n$, $\pm m$) orders forms a circle in momentum space, characterized by a radius of $\sqrt{n^2+m^2}$.
Therefore, for orders possessing identical $\sqrt{n^2+m^2}$ values, they show similar the orientation-dependent reflection coefficient curves as ($\pm n$, $\pm m$) orders, albeit with an angular offset, as illustrated in Figure S7.
Thus, with $P$ set at $\unit[348]{nm}$, the (1,0), (1,1), (2,0), and (2,1) orders with different $\sqrt{n^2+m^2}$ values are selected for investigation as depicted in Figure 4d.
The orientation-dependent curves of these orders differ in both the number of peaks and their positions, revealing the distinct characteristics associated with varying $\sqrt{n^2+m^2}$ values.
These observations offer crucial insights for fine-tuning reflection properties, paving the way for enhanced control over far-field optical responses in polaritonic crystals in anisotropic mediums.

\section{Conclusion}
In conclusion, we investigated the selective excitation of Bloch modes arising from polaritonic crystals with canalized IFCs in a comprehensive way.
By integrating theoretical calculations with experimental analysis, we demonstrate that the polaritons observed in near-field imaging are highly tunable via lattice parameters in both real and momentum spaces, highlighted by the diffractionless polariton propagation. 
Furthermore, we have elucidated how the Bragg resonance orders are delicately adjustable through modifications in the periodicity and lattice orientation of the polaritonic crystals. 
This adjustment notably affects the distribution of the Bloch modes and the magnitude of far-field reflection coefficients.
Intriguingly, we unveiled the potential intricate interplay between the polaritonic crystal lattice and the IFCs, inducing the deviation of theoretical calculations from the experimental results.
Ultimately, our work underscores the vast potential of polaritonic crystal systems in precisely controlling light-matter interactions. 
This opens up promising avenues for the development of reprogrammable polaritonics and advanced nanopolaritonic devices, marking a significant step forward in the manipulation and application of polaritonic phenomena.

\section*{Methods}

\textbf{Sample fabrication and characterization.}
For a twisted bilayer $\upalpha$-MoO$_{\rm 3}$ sample, two flakes were prepared using a two-step exfoliation process. Initially, $\upalpha$-MoO$_{\rm 3}$ flakes were exfoliated from bulk materials provided by SixCarbon Technology, Shenzhen. These flakes were then sequentially stacked, with a specific twist angle, on a SiO$_{2}$ (\unit[285]{nm})/Si substrate, which had been treated with a vacuum plasma cleaner (SUNJUNE PLASMA VP-R3).
Optical microscopy and atomic force microscopy were both utilized to determine the desired thickness of $\upalpha$-MoO$_{\rm 3}$.
To fabricate PoCs within twisted bilayer $\upalpha$-MoO$_{\rm 3}$, electron beam lithography (EBL) followed by reactive ion etching (RIE) was applied to pattern the flakes.

\noindent \textbf{IR \textit{s}-SNOM measurements.}
Near-field images were obtained using a commercially available IR \textit{s}-SNOM system (Bruker nanoIR3s) operating in tapping mode atomic force microscopy (AFM). 
An IR beam with a frequency of \unit[931]{cm$^{-1}$} was generated by a CO$_2$ laser (Access Laser, L4SL-13CO2) and focused on the apex of a gold-coated AFM tip (160AC-GG, OPUS) with tapping frequency $\Omega$ at around \unit[270]{kHz}.
PhPs in $\upalpha$-MoO$_{\rm 3}$ are launched by the tip, propagating and interfering with the reflected wave by the edges of etched holes.
The back-scattered light from the tip was collected with an off-axis parabolic mirror and directed to a HgCdTe (MCT) photodetector to extract the near-field signal from the device. Detected signals were demodulated at the 2nd harmonic to suppress the background.

\section*{Supporting Information}
Definition of flat iso-frequency contour and direction of polariton canalization, calculation of polariton dispersion, calculation of electric field distributions of Bloch modes (PDF).

\section*{Acknowledgements} 
Y.Z.Y. and Z.C.Z. contributed equally to this work. Y.Z.Y., Z.C.Z. and T.J. acknowledge support from the National Natural Science Foundation of China (62175188, 62475194) and the Science and Technology Commission of Shanghai Municipality (23ZR1465800, 23190712300).
X.C. acknowledges support from the National Natural Science Foundation of China (61925504). 
Z.W. acknowledges support from the National Natural Science Foundation of China (62192770, 62192772).
D.H. acknowledges support from the National Natural Science Foundation of China (62305249).
Q.O. acknowledges support from the National Natural Science Foundation of China (52402166), and the Science and Technology Development Fund, Macau SAR (0116/2022/A3, 0065/2023/AFJ).
X.N. acknowledges the support from Research Startup Fund of Central South University (Grant No. 11400-506030109).
The authors acknowledge Renkang Song, Ziheng Shen, Wenhao Su, and Ke Yu for their support in sample preparation and Tianshu Jiang for his valuable discussions.

\section*{Conflict of Interest}
The authors declare no conflict of interest.
\bibliography{Manuscript}

\end{document}